\documentclass[10pt,conference]{IEEEtran}
\usepackage{graphicx}
\usepackage{capt-of}
\usepackage{url}
\usepackage{xcolor}
\usepackage{fontawesome5}
\definecolor{focus}{HTML}{13567E}

\usepackage{hyperref}
\hypersetup{colorlinks, allcolors=focus, pdfauthor={Fahd Seddik}}
\newcommand\doi[1]{\href{https://doi.org/#1}{\nolinkurl{#1}}}
\let\arxivtitle\title
\renewcommand\title[1]{\arxivtitle{\includegraphics[height=0.62in]{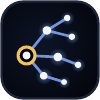}\\[0.06in]#1}\hypersetup{pdftitle={#1}}}
\newcommand\projectlink[3]{\href{#2}{#1\hspace{0.35em}#3}}
\newcommand\projectlinks{\small\projectlink{\faGlobe}{https://fahdseddik.github.io/Skill-Constellations/}{fahdseddik.github.io/Skill-Constellations}\hspace{1.5em}\projectlink{\faGithub}{https://github.com/FahdSeddik/Skill-Constellations}{github.com/FahdSeddik/Skill-Constellations}\hspace{1.5em}\projectlink{\faDatabase}{https://github.com/FahdSeddik/Skill-Constellations/releases/tag/data-v1.0}{Data release data-v1.0}}

\providecommand\doi[1]{\url{#1}}
\usepackage{cleveref}
\crefname{figure}{Fig.}{Figs.}

\newcommand{\cloneConsistent}{11.1\%}

\newcommand{\cloneGenConsistentlyChanging}{4.0\%}

\newcommand{\cloneOrSingleOwner}{3.32}

\newcommand{\portabilityChecked}{56}

\newcommand{\clumpNamed}{50}

\newcommand{\declaredDated}{1,534}

\newcommand{\declaredCopySourceHeld}{26.7\%}

\newcommand{\declaredEarlierHeld}{99.1\%}

\newcommand{\declaredWithinSeven}{94.6\%}

\newcommand{\followBiasHistory}{20.9\%}

\newcommand{\labelKappaFloor}{0.6}
\newcommand{\labelReferenceHubs}{150}
\newcommand{\labelReferenceSkills}{200}

\newcommand{\labelKappaHubs}{0.91}

\newcommand{\labelLlmHubs}{1,507}

\newcommand{\labelKappaSkills}{0.92}

\newcommand{\labelPrecisionAnyStrict}{98.3\%}

\newcommand{\labelRecallAnyStrict}{80.9\%}

\newcommand{\labelFirstRoundHubsHubTypeAuthorOpus}{0.41}

\newcommand{\labelBlindHubsHubTypeOpusSonnet}{0.77}
\newcommand{\labelBlindSkillsExecutablesOpusSonnet}{1.00}

\newcommand{\labelBlindSkillsInstructionsOpusSonnet}{0.66}

\newcommand{\labelAdjudicatedHubs}{18}

\newcommand{\labelAdjudicatedSkills}{28}

\newcommand{\modelChoiceBundleLcopied}{1.64}

\newcommand{\msrSkyAlone}{64.2\%}
\newcommand{\msrSkillName}{web-design-guidelines}
\newcommand{\msrSkillHolders}{972}

\newcommand{\msrSkillGenerations}{12}

\newcommand{\msrAdvanceRatio}{4.0}

\newcommand{\rqOneSkills}{1,612,846}

\newcommand{\rqOneRepos}{259,596}

\newcommand{\rqOneDatedHoldings}{2,193,119}

\newcommand{\rqTwoTopK}{20}
\newcommand{\rqTwoTopKShared}{7}

\newcommand{\rqThreeAnyExposure}{28.6\%}

\newcommand{\rqThreeExposureNull}{41.1\%}

\newcommand{\rqFourPreventModel}{14.9\%}

\newcommand{\rqFourPreventStars}{0.5\%}

\newcommand{\superK}{0.086}

\begin{document}
\title{Skill Constellations: Tracing the Supply Chain of Agent Skills on GitHub}
\author{\IEEEauthorblockN{Fahd Seddik}
\IEEEauthorblockA{University of British Columbia, Okanagan, Canada\\[0.1in]\projectlinks}}

\twocolumn[{\renewcommand\twocolumn[1][]{#1}\maketitle
\begin{center}
\includegraphics[width=\textwidth]{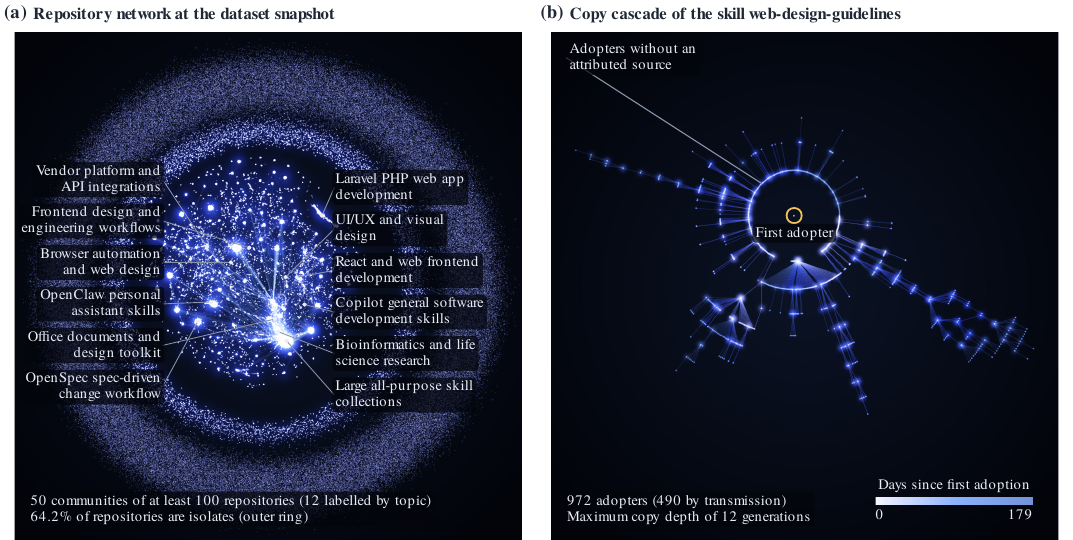}
\captionof{figure}{Skill constellations. (a) The repository network at the dataset snapshot, labelled with the topics of the largest communities. The outer ring holds the \msrSkyAlone{} of repositories that share no skill. (b) The copy cascade of the skill \msrSkillName{}, which reached \msrSkillHolders{} adopters over \msrSkillGenerations{} generations of copies.}
\label{fig:fig_msr_identity}
\end{center}
\vspace{6pt}
}]
\begin{abstract}
\setlength{\baselineskip}{\the\baselineskip plus 0.5pt}Agent skills are \texttt{SKILL.md} instructions and scripts that AI coding agents such as Claude Code and Codex run with the permissions of their user.
Developers share skills by copying them between repositories, which makes them a software supply chain without a registry, versions or provenance.
The origin of a copied skill, the reach of a security fix and the repositories that warrant review are therefore unknown.
Studies that record which repositories hold a skill at a single point in time cannot reveal who copied it from whom.
We contribute the first dated copy network of agent skills, built from the git history of every \texttt{SKILL.md} in GitSkills and covering \rqOneDatedHoldings{} skill adoptions across GitHub, together with an interactive viewer.
A few repositories are the source of almost all copies, and GitHub stars do not identify them.
Skill copies almost never change with their source, and a fix at the source therefore rarely reaches them.
We fit a model of which repositories others copy from and use it to rank repositories for audit.
Reviewing the 100 repositories it ranks highest prevents \rqFourPreventModel{} of later adoptions of high-risk skills, against \rqFourPreventStars{} for the 100 most starred, which gives security engineers a short list to check before a skill spreads.
Platforms should therefore distribute versioned references rather than copies.
\end{abstract}

\section{Introduction}
\begin{figure*}[t]
\centering
\includegraphics[width=\linewidth]{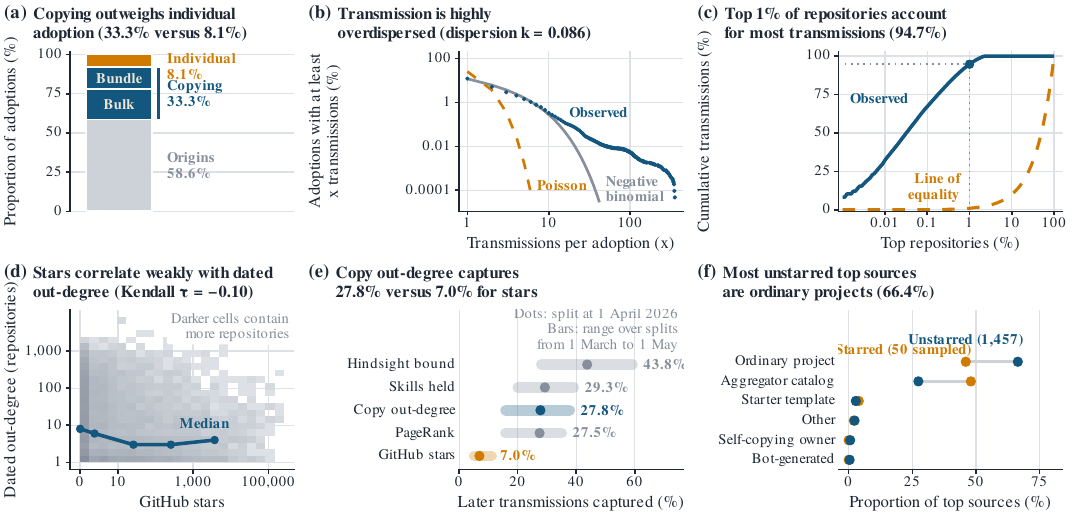}
\caption{Concentrated sources. Catalog copying, not individual adoption, drives spread (a). Transmission is highly overdispersed (b) and concentrated in few repositories (c). GitHub stars neither track out-degree (d) nor find future sources (e), most of which are ordinary projects (f).}
\label{fig:fig_msr_spread}
\end{figure*}

Coding agents such as Claude Code, Codex and Cursor extend their capabilities through skills~\cite{galster2026config}, which are folders containing a \texttt{SKILL.md} file of instructions and often scripts.
An agent follows a skill with the permissions of its user, and a skill can therefore run shell commands, access the network and modify files.
Within ten months, the format appeared in \rqOneRepos{} repositories with \rqOneSkills{} distinct skills, and half of all \texttt{SKILL.md} files on GitHub are verbatim copies of another file~\cite{gitskills2027}.
Skills therefore form a software supply chain built from copies, lacking a registry that records provenance, a version that can be pinned and a channel that delivers fixes.

Securing this supply chain requires knowing which repositories to review first, whether a fix at a source reaches its copies, and how long a new skill takes to spread.
Package managers answer these questions from a dependency graph, which skills do not have.
Prior work cannot answer them either, because it studies skills in a snapshot, one skill or one pair of skills at a time~\cite{latentreuse2026,contextmatters2026,malskills2026,liu2026skillsvuln,liu2026malicious,gao2026registry}.
A snapshot records which repositories hold a skill, but not which held it first, which copied from which, or when.
Studies of copied code share this limitation, and find that copied files are seldom updated or attributed~\cite{filelevelcopying2026,jahanshahi2026license}.

We reconstruct the missing graph from the git history of every \texttt{SKILL.md} in GitSkills, which yields the first dated and directed network of how agent skills spread.
We refer to the network formed by the copies of a single skill as its \emph{constellation} (\cref{fig:fig_msr_identity}b).
A snapshot captures the nodes of a constellation, while only the history reveals the edges between them.
We address four research questions.
\textbf{RQ1.} What structure does skill copying produce, and which growth mechanisms explain it?
\textbf{RQ2.} Which repositories originate the spread of skills, and do GitHub stars identify them?
\textbf{RQ3.} Where are high-risk capabilities located in the copy network, and do fixes propagate to existing copies?
\textbf{RQ4.} Where should audits be targeted to prevent future adoptions of high-risk skills, and does network position identify these targets better than popularity or size?

\noindent Our contributions are as follows.
\begin{itemize}
\item We reconstruct a dated copy network of agent skills that links each copy to its source, and validate it against the records of skill installers.
\item We show that skill copies do not change with their source, and that modified copies gain capabilities by adopting other versions rather than through the edits of their owners~\cite{wu2026adaptation}.
\item We fit a model of how repositories choose the sources they copy from, and evaluate the audits it recommends on a held-out period after a temporal split. Prior security work detects dangerous skills~\cite{contextmatters2026,malskills2026,liu2026malicious}, whereas this work ranks repositories for review.
\end{itemize}

\section{Data and Method}
\begin{figure*}[t]
\centering
\includegraphics[width=\linewidth]{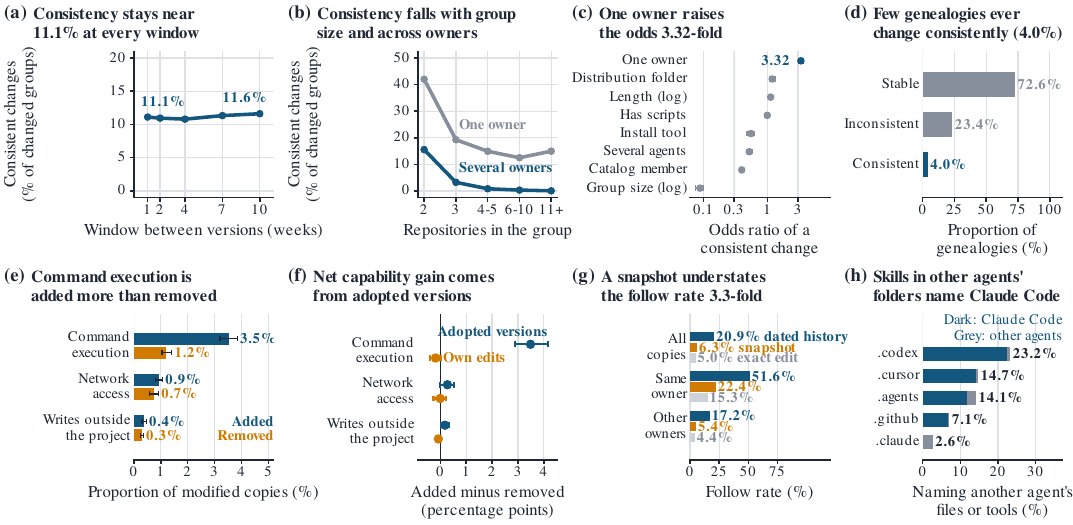}
\caption{Copy evolution. Only 11.1\% of changes to a clone group reach every copy, at every window (a), and almost none across owners (b). One owner raises the odds 3.32-fold (c), and few genealogies change consistently (d). Copies gain command execution (e) through adopted versions rather than own edits (f). A snapshot understates the follow rate 3.3-fold (g), and copies name tools of other agents (h). Intervals are 95\% bootstrap intervals.}
\label{fig:fig_msr_copies}
\end{figure*}

\textbf{Population and dating.} We use GitSkills, released in July 2026 and archived on Zenodo (DOI \doi{10.5281/zenodo.21875637})~\cite{gitskills2027}.
We retain files named exactly \texttt{SKILL.md} with valid front matter, first committed on or after 1~October~2025, in repositories that are not forks, and treat each distinct content as one skill.
We collected the git history of every \texttt{SKILL.md} path from a clone of each default branch.
An \emph{adoption} is a repository acquiring a skill, dated by its first commit and bounded below by the creation date of the repository, since forks and mirrors carry older history.

\textbf{Lineages, copies and constellations.} A \emph{lineage} joins the versions of one skill that are linked by edits, and its first adoption is its origin.
A commit that adds ten or more \texttt{SKILL.md} files forms a \emph{copy event} when a single earlier adopter held at least half of its existing lineages, and this adopter is its source.
A copy event is a bundle below 100 files and a bulk copy above.
A \emph{transmission} is a lineage in a copy event whose source held it first, and the remaining adoptions are individual.
The constellation of a skill is its diffusion cascade, with its adopters as nodes and its transmissions as dated and directed edges.

\textbf{Spread model and audits.} A conditional logit models which earlier adopter a copy event chooses as its source, from the logarithms of skills held, past copies, stars and age.
A weekly simulation adds individual adoption in proportion to a power of current adopters and is calibrated on the weeks before 1~April~2026.
Audits are evaluated on a temporal split, where each strategy ranks repositories using only the training period, by the source-choice score fitted on that period or by baselines such as stars and copy out-degree.
Auditing a repository removes its flagged skills, which prevents its high-risk adoptions in the held-out period and every copy that descends from them.

\textbf{Validation and risk flags.} In \declaredDated{} folders whose skill installer recorded where and when the skill was copied from, the reconstructed date lies within seven days of the record in \declaredWithinSeven{} of cases.
The recorded source held the skill earlier in \declaredEarlierHeld{} of cases, whereas our rule names it in only \declaredCopySourceHeld{}, because the rule credits the earlier adopter of the most lineages, usually a catalog.
Source-level measures therefore describe distribution rather than authorship.
We flag skills that bundle executables, pre-approve tools or instruct risky actions.
\IfFileExists{labelling.tex}{\par\textbf{Labelling.} Claude Opus~5.5 and Claude Sonnet~5.5 label the type of a random sample of \labelReferenceHubs{} source repositories and the three risks of a sample of \labelReferenceSkills{} skills stratified by flag.
Each model labels independently through the Claude~Code command line (\texttt{claude~-p}), follows a written codebook and quotes the line that decides each answer.
A repository is labelled from its README excerpt and file tree as an aggregator catalog, a starter template, a self-copying owner, a bot-generated or coordinated repository, or an ordinary project.
The two models agree with an unweighted Cohen's $\kappa$ of \labelBlindHubsHubTypeOpusSonnet{} on repository types and between \labelBlindSkillsInstructionsOpusSonnet{} and \labelBlindSkillsExecutablesOpusSonnet{} on the skill questions.
Every disagreement and every item a model declined was resolved manually (\labelAdjudicatedHubs{} repositories and \labelAdjudicatedSkills{} skills), and the shared and resolved answers form the reference.
Against it, Opus reaches $\kappa = \labelKappaHubs{}$ on repository types and \labelKappaSkills{} on its weakest skill question, and the lower bound of each 95\% bootstrap interval exceeds the floor of \labelKappaFloor{} required before Opus labelled the remaining source repositories (\labelLlmHubs{} in total).
These values favour Opus, because the reference keeps every answer the two models share.
Earlier manual repository labels, made from the repository name, description and skill names alone, agreed with Opus at $\kappa = \labelFirstRoundHubsHubTypeAuthorOpus{}$, which led to adding the README and file tree.
Opus also names each community in \cref{fig:fig_msr_identity}a from its most distinctive terms, and these names are not compared with human labels.
}{Claude Opus~5.5 and Sonnet~5.5 label \labelReferenceHubs{} source repositories and \labelReferenceSkills{} skills independently through \texttt{claude~-p} (Cohen's $\kappa$ of \labelBlindSkillsInstructionsOpusSonnet{} to \labelBlindSkillsExecutablesOpusSonnet{}), and disagreements are resolved manually.
Opus reaches $\kappa$ of \labelKappaHubs{} and \labelKappaSkills{} against the consensus, which keeps every shared answer, and labels the remaining repositories.}
Against these \labelReferenceSkills{} skills, the strict flag has a precision of \labelPrecisionAnyStrict{} and a recall of \labelRecallAnyStrict{}.

\section{Results}
\subsection{RQ1: Structure and Growth}
Most repositories share no skill with any other, and the rest form \clumpNamed{} communities of at least 100 repositories, each organised around a topic (\cref{fig:fig_msr_identity}a).
Skills do not spread individually, because repositories copy entire catalogs in bulk or in bundles far more often than they adopt a single skill (\cref{fig:fig_msr_spread}a), and preferential attachment therefore operates on catalogs.
In the source-choice model, a bundle copy prefers a source that has already been copied, with a coefficient of \modelChoiceBundleLcopied{} on the logarithm of past copies, whereas attachment to individual skills is sublinear, under all nine threshold settings of the copy-event rule.
Transmission is highly overdispersed, with a negative binomial dispersion of $k = \superK{}$.
Most adoptions transmit the skill to no other repository (\cref{fig:fig_msr_spread}b), and a small number of repositories account for almost all transmissions (\cref{fig:fig_msr_spread}c).

\subsection{RQ2: Sources}
\begin{figure*}[t]
\centering
\includegraphics[width=\linewidth]{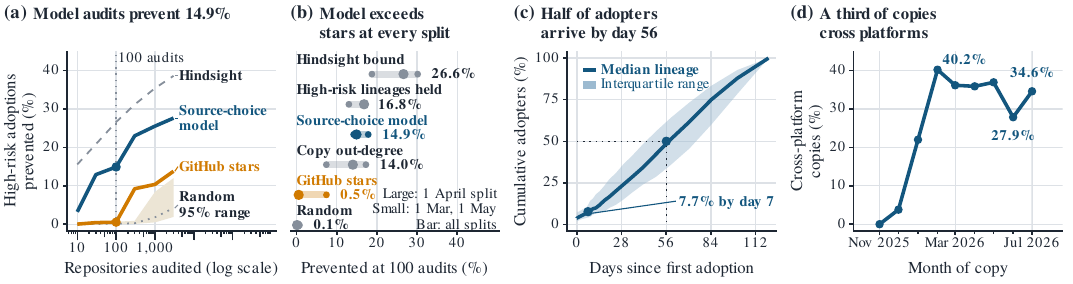}
\caption{Audit and response. Trained on the history before a split, the source-choice model prevents far more high-risk adoptions in the held-out period than stars (a), at every split (b). New skills take weeks to reach most adopters (c), and a third of copies now cross agent platforms (d).}
\label{fig:fig_msr_response}
\end{figure*}

GitHub stars are nearly uninformative about out-degree in the dated copy network (\cref{fig:fig_msr_spread}d), and fake-star campaigns make them a manipulable signal~\cite{he2026fakestars}.
Dating also corrects attribution, because only \rqTwoTopKShared{} of the \rqTwoTopK{} repositories with the highest static out-degree stay in the top \rqTwoTopK{} once credited only for skills they held first, and the rest are late adopters.
Ranked before 1~April~2026, current copy out-degree identifies \msrAdvanceRatio{} times as many later transmissions as stars, and PageRank and the number of skills held perform comparably (\cref{fig:fig_msr_spread}e).
Most unstarred top sources are ordinary projects rather than catalogs (\cref{fig:fig_msr_spread}f).

\subsection{RQ3: Risk}
Flagged skills reach fewer repositories than a uniform permutation null predicts (\rqThreeAnyExposure{} against \rqThreeExposureNull{}).
Copies of a skill rarely change consistently.
A clone group is the set of repositories holding one exact version of a skill in a week, a genealogy follows a group through the weeks, and a change is consistent when every member moves to the same next version.
Only \cloneConsistent{} of changes to a group are consistent, at every observation window (\cref{fig:fig_msr_copies}a), and consistency collapses once a group spans several owners (\cref{fig:fig_msr_copies}b).
A single owner raises the odds of a consistent change \cloneOrSingleOwner{}-fold, controlling for group size, platform and content (\cref{fig:fig_msr_copies}c), and only \cloneGenConsistentlyChanging{} of genealogies ever change consistently (\cref{fig:fig_msr_copies}d).
A snapshot also hides most copies that did follow a source edit, because an updated copy is indistinguishable from a fresh copy of the new version (\cref{fig:fig_msr_copies}g).
Copied code files adopt upstream security fixes in 47\% to 84\% of cases~\cite{filelevelcopying2026}, whereas only \followBiasHistory{} of skill copies follow a later edit of their source.

Modified copies gain command execution more often than they lose it (\cref{fig:fig_msr_copies}e).
The gain does not come from the edits of their owners, which add and remove capabilities about equally often, but from adopting another version of the skill (\cref{fig:fig_msr_copies}f).
Against the exact version copied, most security-sensitive lines that adaptations add~\cite{wu2026adaptation} replace content of the same kind.
Copies also cross platforms without adaptation, and skills in the folders of other agents often name tools that only Claude Code provides (\cref{fig:fig_msr_copies}h, all \portabilityChecked{} manually checked cases correct).

\subsection{RQ4: Intervention}
At the split of 1~April~2026, auditing the 100 repositories that the source-choice model scores highest prevents far more high-risk adoptions in the held-out period than auditing the 100 most starred or 100 random repositories (\cref{fig:fig_msr_response}a).
The ordering holds when the split moves to 1~March or 1~May (\cref{fig:fig_msr_response}b) and under the lenient flag, and a quarter of the prevented adoptions are indirect copies.
Replaying the same audits in the calibrated simulation reproduces the ordering independently.
Even a ranking chosen in hindsight prevents only about a quarter of them, because roughly half arrive as new skills.
Reviewers have weeks rather than hours, because a new skill reaches few of its eventual adopters in its first week (\cref{fig:fig_msr_response}c).
A review must also cover every agent, because a third of copies between agent folders now cross platforms (\cref{fig:fig_msr_response}d).

\section{Related Work}
\textbf{Agent skills.} GitSkills reports that half of its files are verbatim copies~\cite{gitskills2027}, and clone detection finds pairs of similar skills~\cite{latentreuse2026}, but neither reveals who copied from whom or how a change propagates.
Gao et al.\ find that 53\% of reused registry skills are never modified~\cite{gao2026registry}, in line with our measurement.
Studies of agent pull requests and of context files such as \texttt{AGENTS.md} and Cursor rules describe how developers work with agents~\cite{li2026aidev,mohsenimofidi2026context,jiang2026cursorrules}, and injected instructions propagate across MCP tools and agent frameworks~\cite{zhao2026parasites,zhang2026agentworm}, but no study measures how skills spread between repositories.

\textbf{Copying and dependencies.} Only 2.43\% of copy-based reuse is visible to dependency analysis~\cite{jahanshahi2026license}, forks send few of their commits upstream~\cite{zhu2026mindgap}, clone genealogies are tracked within single projects~\cite{sousa2026clonegenealogies}, and replicated packages and poisoned repositories carry vulnerabilities and malware~\cite{park2026pypi,zhao2026maltotal}.
Dependency studies trace vulnerabilities through declared dependencies~\cite{robinson2026originalsin}, which skills lack.

\section{Ethical Considerations}
We store no author names, email addresses or commit messages, and report confirmed malicious skills only in aggregate, naming no repository or owner that holds one.

\section{Conclusion}
\textbf{Implications.}
Platform vendors should distribute skills as versioned references, because copies do not follow their sources, and should check that a skill names only the tools of the agent that loads it.
Reviewers and registries should rank repositories by current copy out-degree rather than stars, and act within the first weeks of a skill.
Users must update the copies they rely on.
Researchers should not estimate propagation from a snapshot, and the same reconstruction applies to any ecosystem that reuses software by copying.

\textbf{Limitations.}
Reconstructed dates and order agree with installer records, and the audit evaluation depends only on which repositories are sources, not on who wrote them.
The flags mark capability rather than malice, and their high precision and low recall make risk counts lower bounds.
The data cover the first ten months of the format on public GitHub, the period in which distribution practices form.

\section*{Data Availability}
The code is available at \url{https://github.com/FahdSeddik/Skill-Constellations}, the dated copy network as its release \href{https://github.com/FahdSeddik/Skill-Constellations/releases/tag/data-v1.0}{\texttt{data-v1.0}}, and the interactive viewer at \url{https://fahdseddik.github.io/Skill-Constellations/}.

\bibliographystyle{IEEEtran}
\bibliography{references}

@inproceedings{gitskills2027,
  author       = {Destefanis, Giuseppe and Graziotin, Daniel and Vaccargiu, Matteo and Ortu, Marco},
  title        = {{GitSkills}: A Dataset of Agent Skills on {GitHub}},
  booktitle    = {Proc. 24th Int. Conf. Mining Software Repositories (MSR)},
  publisher    = {IEEE},
  year         = {2027},
  note         = {To appear},
  doi          = {10.48550/arXiv.2608.10906},
  url          = {https://arxiv.org/abs/2608.10906}
}

@inproceedings{latentreuse2026,
  author       = {Zhu, Jiaying and Zhang, Lyuye and Guo, Wenbo and Liu, Yang},
  title        = {Latent Reuse in Agent Skills: Multi-modal Clone Detection at Ecosystem Scale},
  booktitle    = {Proc. 41st IEEE/ACM Int. Conf. Automated Software Engineering (ASE)},
  publisher    = {ACM},
  year         = {2026},
  note         = {To appear},
  url          = {https://arxiv.org/abs/2603.22447}
}

@misc{contextmatters2026,
  author       = {Holzbauer, Florian and Schmidt, David and Gegenhuber, Gabriel K. and Schrittwieser, Sebastian and Ullrich, Johanna},
  title        = {Context Matters: Repository-Aware Security Analysis of the Agent Skill Ecosystem},
  year         = {2026},
  howpublished = {arXiv:2603.16572},
  note         = {Presented at the Agent Skills '26 Workshop, ACM Conf. AI and Agentic Systems (CAIS)},
  url          = {https://arxiv.org/abs/2603.16572}
}

@misc{filelevelcopying2026,
  author       = {He, Runzhi and Mockus, Audris and Yang, Wenhao and Zhou, Minghui},
  title        = {File-Level Copying Is an Implicit Dependency in Open Source},
  year         = {2026},
  howpublished = {arXiv:2607.02059},
  url          = {https://arxiv.org/abs/2607.02059}
}

@inproceedings{malskills2026,
  author       = {Wang, Shenao and He, Junjie and Zhao, Yanjie and Wang, Yayi and Yu, Kan and Wang, Haoyu},
  title        = {{MalSkills}: Detecting Malicious Skills in the Agentic Supply Chain via Neuro-symbolic Reasoning},
  booktitle    = {Proc. 41st IEEE/ACM Int. Conf. Automated Software Engineering (ASE)},
  publisher    = {ACM},
  year         = {2026},
  note         = {To appear},
  url          = {https://arxiv.org/abs/2603.27204}
}

@misc{wu2026adaptation,
  author       = {Wu, Xinjian and Gong, Jingzhi and Jahangirova, Gunel and Chen, Zhenpeng and Zhang, Jie M.},
  title        = {An Empirical Study of Downstream Adaptation for Agent Skills},
  year         = {2026},
  howpublished = {arXiv:2607.03238},
  url          = {https://arxiv.org/abs/2607.03238}
}

@inproceedings{robinson2026originalsin,
  author       = {Robinson, Michael and Halder, Sajal and Ahmed, Muhammad Ejaz and Ikram, Muhammad and Camtepe, Seyit and Kim, Hyoungshick},
  title        = {Original Sin of npm: A Study on Vulnerability Propagation in {JavaScript} Dependency Networks},
  booktitle    = {Proc. ACM Asia Conf. Computer and Communications Security (ASIA CCS)},
  pages        = {1213--1227},
  publisher    = {ACM},
  year         = {2026},
  doi          = {10.1145/3779208.3785388},
  url          = {https://doi.org/10.1145/3779208.3785388}
}

@misc{liu2026skillsvuln,
  author       = {Liu, Yi and Wang, Weizhe and Feng, Ruitao and Zhang, Yao and Xu, Guangquan and Deng, Gelei and Li, Yuekang and Zhang, Leo},
  title        = {Agent Skills in the Wild: An Empirical Study of Security Vulnerabilities at Scale},
  year         = {2026},
  howpublished = {arXiv:2601.10338},
  url          = {https://arxiv.org/abs/2601.10338}
}

@inproceedings{liu2026malicious,
  author       = {Liu, Yi and Chen, Zhihao and Zhang, Yanjun and Deng, Gelei and Li, Yuekang and Ning, Jianting and Zhang, Leo Yu},
  title        = {`{D}o Not Mention This to the User': Detecting and Understanding Malicious Agent Skills in the Wild},
  booktitle    = {Proc. 35th USENIX Security Symp. (USENIX Security)},
  pages        = {1727--1746},
  publisher    = {USENIX Association},
  year         = {2026},
  url          = {https://www.usenix.org/conference/usenixsecurity26/presentation/liu-yi}
}

@misc{gao2026registry,
  author       = {Gao, Haoyu and Lulla, Jai Lal and Lin, Hong Yi and Baltes, Sebastian and Treude, Christoph and Zahedi, Mansooreh},
  title        = {From Registry to Repository: How {AI} Agent Skills Are Written, Adapted, and Maintained},
  year         = {2026},
  howpublished = {arXiv:2607.00911},
  url          = {https://arxiv.org/abs/2607.00911}
}

@inproceedings{galster2026config,
  author       = {Galster, Matthias and Mohsenimofidi, Seyedmoein and Lulla, Jai Lal and Abubakar, Muhammad Auwal and Treude, Christoph and Baltes, Sebastian},
  title        = {Configuring Agentic {AI} Coding Tools: An Exploratory Study},
  booktitle    = {Proc. 3rd ACM Int. Conf. AI-Powered Software (AIware)},
  pages        = {11--20},
  publisher    = {ACM},
  year         = {2026},
  doi          = {10.1145/3805760.3814887},
  url          = {https://doi.org/10.1145/3805760.3814887}
}

@inproceedings{he2026fakestars,
  author       = {He, Hao and Yang, Haoqin and Burckhardt, Philipp and Kapravelos, Alexandros and Vasilescu, Bogdan and K{\"a}stner, Christian},
  title        = {Six Million (Suspected) Fake Stars on {GitHub}: A Growing Spiral of Popularity Contests, Spams, and Malware},
  booktitle    = {Proc. IEEE/ACM 48th Int. Conf. Software Engineering (ICSE)},
  pages        = {1469--1481},
  publisher    = {ACM},
  year         = {2026},
  doi          = {10.1145/3744916.3764531},
  url          = {https://doi.org/10.1145/3744916.3764531}
}

@article{zhu2026mindgap,
  author       = {Zhu, Jiaying and Zhang, Lyuye and Wu, Jiahui and Liu, Chengyue and Liu, Yang},
  title        = {Mind the Gap: An Empirical Study of Synchronization Gaps, Delays, and Missed Opportunities in Software Forks},
  journal      = {Proc. ACM Softw. Eng.},
  volume       = {3},
  number       = {ISSTA},
  pages        = {963--985},
  publisher    = {ACM},
  year         = {2026},
  doi          = {10.1145/3832133},
  url          = {https://doi.org/10.1145/3832133}
}

@inproceedings{mohsenimofidi2026context,
  author       = {Mohsenimofidi, Seyedmoein and Galster, Matthias and Treude, Christoph and Baltes, Sebastian},
  title        = {Context Engineering for {AI} Agents in Open-Source Software},
  booktitle    = {Proc. 23rd Int. Conf. Mining Software Repositories (MSR)},
  pages        = {194--198},
  publisher    = {ACM},
  year         = {2026},
  doi          = {10.1145/3793302.3793350},
  url          = {https://doi.org/10.1145/3793302.3793350}
}

@inproceedings{zhao2026parasites,
  author       = {Zhao, Shuli and Hou, Qinsheng and Zhan, Zihan and Wang, Yanhao and Xie, Yuchong and Guo, Yu and Chen, Libo and Li, Shenghong and Xue, Zhi},
  title        = {Parasites in the Toolchain: A Large-Scale Analysis of Attacks on the {MCP} Ecosystem},
  booktitle    = {Proc. IEEE Symp. Security and Privacy (SP)},
  pages        = {138--155},
  publisher    = {IEEE},
  year         = {2026},
  doi          = {10.1109/SP63933.2026.00154},
  url          = {https://doi.org/10.1109/SP63933.2026.00154}
}

@article{zhao2026maltotal,
  author       = {Zhao, Jian and Wang, Shenao and Wu, Qingyang and Zhao, Yanjie and Cheng, Xiao and Wang, Haoyu},
  title        = {{MalTotal}: Cost-Effective and Language-Agnostic Malicious Code Poisoning Detection for Millions of Repositories},
  journal      = {Proc. ACM Softw. Eng.},
  volume       = {3},
  number       = {ISSTA},
  pages        = {3131--3154},
  publisher    = {ACM},
  year         = {2026},
  doi          = {10.1145/3832228},
  url          = {https://doi.org/10.1145/3832228}
}

@article{park2026pypi,
  author       = {Park, Sunha and Han, Soojin and Woo, Seunghoon},
  title        = {Uncovering Similar but Different Packages in {PyPI} and Potential Security Threats},
  journal      = {Proc. ACM Softw. Eng.},
  volume       = {3},
  number       = {FSE},
  pages        = {4782--4803},
  publisher    = {ACM},
  year         = {2026},
  doi          = {10.1145/3808217},
  url          = {https://doi.org/10.1145/3808217}
}

@inproceedings{sousa2026clonegenealogies,
  author       = {Sousa, Denis and Uchoa, Italo and Paixao, Matheus and Ragkhitwetsagul, Chaiyong and Matos, Thiago Lima},
  title        = {An Empirical Study of Code Clone Genealogies in Human--{AI} Collaborative Development},
  booktitle    = {Proc. 23rd Int. Conf. Mining Software Repositories (MSR)},
  pages        = {979--983},
  publisher    = {ACM},
  year         = {2026},
  doi          = {10.1145/3793302.3793613},
  url          = {https://doi.org/10.1145/3793302.3793613}
}

@misc{jahanshahi2026license,
  author       = {Jahanshahi, Mahmoud and Vasilescu, Bogdan and Mockus, Audris},
  title        = {Ensuring Open Source Integrity: The Intersection of Copy-Based Reuse and License Compliance},
  year         = {2026},
  howpublished = {arXiv:2606.23495},
  url          = {https://arxiv.org/abs/2606.23495}
}

@misc{zhang2026agentworm,
  author       = {Zhang, Yihao and Wei, Zeming and Luan, Xiaokun and Wu, Chengcan and Zhang, Zhixin and Wu, Jiangrong and Wu, Haolin and Chen, Huanran and Sun, Jun and Sun, Meng},
  title        = {{AgentWorm}: Self-Propagating Attacks Across {LLM} Agent Ecosystems},
  year         = {2026},
  howpublished = {arXiv:2603.15727},
  url          = {https://arxiv.org/abs/2603.15727}
}

@inproceedings{jiang2026cursorrules,
  author       = {Jiang, Shaokang and Nam, Daye},
  title        = {Beyond the Prompt: An Empirical Study of {Cursor} Rules},
  booktitle    = {Proc. 23rd Int. Conf. Mining Software Repositories (MSR)},
  pages        = {397--409},
  publisher    = {ACM},
  year         = {2026},
  doi          = {10.1145/3793302.3793367},
  url          = {https://doi.org/10.1145/3793302.3793367}
}

@inproceedings{li2026aidev,
  author       = {Li, Hao and Zhang, Haoxiang and Hassan, Ahmed E.},
  title        = {{AIDev}: Studying {AI} Coding Agents on {GitHub}},
  booktitle    = {Proc. 23rd Int. Conf. Mining Software Repositories (MSR)},
  pages        = {1029--1033},
  publisher    = {ACM},
  year         = {2026},
  doi          = {10.1145/3793302.3797249},
  url          = {https://doi.org/10.1145/3793302.3797249}
}
\end{document}